\documentclass[aps,pra,epsfigure,twocolumn,showpacs]{revtex4}
\usepackage{dcolumn}
\usepackage{bm}
\usepackage{graphicx}
\usepackage{amsmath}
\usepackage{latexsym}
\usepackage{amsfonts}
\usepackage{amssymb}
\usepackage{epsfig}
\usepackage{array}

\newcommand{\ket}[1]{\left\vert#1\right\rangle}
\newcommand{\miniket}[1]{\vert#1\rangle}
\newcommand{\modul}[1]{\left\vert#1\right\vert}

\newcommand{\pro}[3]{\left\vert#1\rangle_{#2}\langle#3\right\vert}

\newcommand{\proj}[2]{\left\vert#1\rangle\langle#2\right\vert}
\newcommand{\miniproj}[2]{\vert#1\rangle\langle#2\vert}
\newcommand{\bra}[1]{\left\langle#1\right\vert}

\begin{document}
\title{Noise resilience and entanglement evolution in two non-equivalent classes of quantum algorithms}

\author{C. Di Franco, M. Paternostro, and M. S. Kim}

\affiliation{School of Mathematics and Physics, Queen's University, Belfast BT7
  1NN, United Kingdom}

\begin{abstract}
The speed-up provided by quantum algorithms with respect to their classical counterparts is at the origin of scientific interest in quantum computation. However, the fundamental reasons for such a speed-up are not yet completely understood and deserve further attention. In this context, the classical simulation of quantum algorithms is a useful tool that can help us in gaining insight. Starting from the study of general conditions for classical simulation, we highlight several important differences between two non-equivalent classes of quantum algorithms. We investigate their performance under realistic conditions by quantitatively studying their resilience with respect to static noise. This latter refers to errors affecting the inital preparation of the register used to run an algorithm. We also compare the evolution of the entanglement involved in the different computational processes.
\end{abstract}


\maketitle

\section{Introduction}
\label{intro}

One of the major reasons for investigating quantum computation is the possibility for a quantum processor to outperform any analogous classical device~\cite{nielsenchuang}. The design of genuine quantum algorithms and the study of the reasons for their speed-up with respect to classical counterparts have been the center of considerable interest. It is now generally accepted that, for pure states, quantum correlations spread over a sufficiently large number of elements of a register~\cite{jozsa} and quickly growing with the size of the register itself~\cite{vidal1} are a necessary requirement for the speed-up. However, for registers prepared in mixed states, the requirements are still largely unknown. It has been conjectured that, for mixed states, the criteria mentioned above are not sufficient.

The study of classical simulation of quantum algorithms can help us in understanding the role of inherently quantum phenomena in computational problems. Considerable effort has been made in this direction, with proposals for classical simulations designed in experimental setups ranging from nuclear magnetic resonance~\cite{algoritmiimplementazione}, to cavity-quantum electrodynamics and linear optics~\cite{simulo}. It has been pointed out by Meyer that it is possible to classically simulate quantum algorithms that rely on the use of balanced linear superpositions of the computational states of a register~\cite{meyer}. Such an initial state can then be reinterpreted as the state of a multilevel particle by neglecting the multipartite nature of the register. This allows one to reinterpret quantum entanglement in terms of simple coherences. In this paper, we use the quantum average algorithm~\cite{groveraverage} as the representative of a class of quantum protocols (from now on indicated as {\it non-polylocal}) which is non-equivalent to the class identified by Meyer (labeled as {\it polylocal}). The latter is represented, in our study, by the quantum search algorithm~\cite{groversearch}.

A clear difference between the two classes is the ``nature'' of the initial state of the register. Polylocal algorithms use initially separable states and generate entanglement during their performance~\cite{braunsteinpati}. Differently, non-polylocal protocols exploit entangled initial resources. As a result, in this second class of algorithms, the state of the register cannot be put in correspondence with an unbiased state of a multilevel system.  This prevents the use of general arguments {\it \'a la} Meyer. Moreover, the entangled resource itself is ``consumed'' during the processing of non-polylocal algorithms. It is also interesting to notice a close analogy between these protocols and the measurement-based model for computation~\cite{cluster}.

The different use of entanglement in the two classes of problems makes their quantitative comparison difficult. In our study, we use the influence of noise (and thus the introduction of classical correlations in the algorithms) as a useful tool for the investigation into differences between the representatives of polylocal and non-polylocal algorithms. For the quantum average algorithm, although a fragile GHZ-like state~\cite{ghz} is used, we find a considerable resilience to static noise. Moreover, the GHZ-like nature of the resource remains unchanged, leading us to conclude that this specific form of entanglement plays a crucial role in the performance of the algorithm. 

This paper is organized as follows. In Sec.~\ref{funzionamento} we provide a brief explanation of how the representatives of the two different classes work and briefly comment on their classical simulation. In Sec.~\ref{rumore} we investigate the performances of these protocols in the presence of static noise, focusing our attention on the quantum average algorithm for which, to the best of our knowledge, noise resilience has never been studied. The evolution of entanglement is studied in Sec.~\ref{entanglement}, where a clear picture of the salient properties of the quantum average algorithm is provided. Finally, in Sec.~\ref{conclusioni} we summarize our results.

\section{Dynamics of the representative algorithms}
\label{funzionamento}
\subsection{Polylocal class: Quantum search algorithm}
\label{grovercerca}
The quantum search algorithm~\cite{groversearch} is designed to find a searched item in a randomly ordered database of length {$L$} in an {${\cal O} (\sqrt{L})$} time~\cite{commentonotazione}. If we want to carry out the same search using a classical algorithm, on average {$\frac{L}{2}$} steps are required, as we need to analyze the items one by one until the searched one is found. In order to describe the algorithm, we assume that each item of the database is labeled by a binary number between 0 and {$L-1$} with {$L = 2^n$} and $n$ an integer. Thus, the task becomes the identification of the number labeling the searched item. Using an {$n$}-qubit system, each state of the computational basis corresponds to a binary number in the set {$\{0,...,L-1\} $}. For example, the state {$\ket{0 \cdot \cdot \cdot 0101}$} corresponds to 101. The algorithm consists of an alternating sequence of operators, as a result of which the target state can be found in {${\cal O} (\sqrt{L})$} queries. The initial state of the register is the superposition {$\ket{\tilde{0}} = ({1}/{\sqrt{L}})\sum_{i} {\ket{\underline{i}}} $} with {$\ket{\underline{i}}$} one of the states of the computational basis and {$i=0,...,L-1$}. This state can be obtained by applying {$\hat{H}^{\otimes n}$} to {$\ket{\underline{0}}$} with {$\hat{H}$} the Hadamard transform
$\hat{H} = \frac{1}{\sqrt{2}}
\begin{pmatrix}
1&1\\
1&-1
\end{pmatrix}$
written in the single-qubit basis {$ \{ \ket{0} , \ket{1} \} $}. The sequence of operators, or {\it Grover iterate}, is made out of the following four steps:
\begin{itemize}
\item{Query of the oracle~\cite{oracolo}: The oracle performs a phase-flip on the searched state. This is {\it effectively} performed through the operator {$\hat{\mathbb{I}} - 2 \proj{s}{s}$} with {$\ket{s}$} the searched state and $\hat{{\mathbb I}}$ the $L \times L$ identity matrix.}
\item{Application of {$\hat{H}^{\otimes n}$}.}
\item{Application of {$2\proj{\underline{0}}{\underline{0}} - \hat{\mathbb{I}}$}.}
\item{Application of {$\hat{H}^{\otimes n}$}.}
\end{itemize}
The query of the oracle can be seen as being at the heart of the algorithm. For a generic input state, the oracle marks the searched state flipping the sign of its amplitude. Classically, this corresponds to a checking function which produces a different output depending on whether the input is the searched one or not.
The last three operations can be seen as the application of the operator {$2\proj{\tilde{0}}{\tilde{0}}-\hat{\mathbb{I}}$}, which carries out an inversion about the mean. 
By iterating this sequence, the state of the {$n$}-qubit system oscillates between the equally weighted superposition {$\ket{\tilde{0}}$} and the searched state. The first maximum of this oscillation is obtained after {$R$} iterations of the sequence, where {$R$} is the closest integer to the real number
\begin{equation}
X = \frac{\arccos \, \sqrt{\frac{1}{L}}}{2 \, \arccos \, \sqrt{\frac{L-1}{L}}} .
\end{equation}
For large {$L$}, {$X$} (and therefore {$R$}) becomes {${\cal O} (\sqrt{L})$}, so that the number of steps required to find the state {$\ket{s}$} with almost certainty scales as {$\sqrt{L}$}.

\subsection{Non-polylocal class: Quantum average algorithm}

Suppose we have {$N$} values {$\nu_j \in [-1,1]$}. A well-known computer-science problem is to find the order of magnitude of the average {$\mu$}, defined by {$\mu=\frac{1}{N} \sum_{j=1}^{N} \nu_j$}. Classically, if we pick up {$m$} random samples from the {$N$}-value set, the average evaluated out of them will be distributed according to a Gaussian centered at the actual average and with standard deviation {${\cal O} (\frac{1}{\sqrt{m}})$}, as obtained in virtue of the central limit theorem. This means that, with high probability, the estimated average lies within {${\cal O} (\frac{1}{\sqrt{m}})$ of the true average. Suppose that, for a specific problem to be solved, {$\mu$} must be known with a precision at least equal to {$\epsilon$}. This means that {${\cal O} (\frac{1}{\sqrt{m}}) \le \epsilon$} and thus {$m \ge \Omega (\frac{1}{\epsilon^2})$} samples are needed to estimate the average with a precision of {$\epsilon$}. If we want to know the order of magnitude of the average, we need {$\Omega (\frac{1}{\mu^2})$} samples. A speed-up in finding the solution to this problem can be achived by using the following quantum algorithm, which is able to estimate the ratio {$\frac{\modul{\mu}}{\theta}$}, for a fixed {$\theta>0$}, in a number of steps independent of {$\mu$} and {$\theta$} and depending only on the precision we want to have with respect to the estimate of this ratio.
\begin{figure}[t]
\psfig{figure=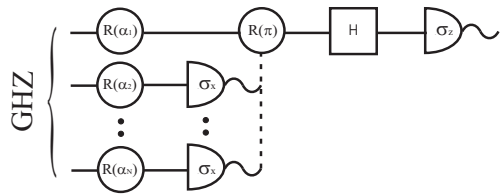,width=6.5cm,height=2.5cm}
\caption{Logical circuit for the quantum average algorithm~\cite{groveraverage}. The input is the GHZ state in Eq.~(\ref{input}). Single-qubit rotations are indicated as $R(\alpha_j)$ with $\alpha_j=\frac{\nu_j}{N\theta}$ and the Hadamard gate $\hat{H}$ is also shown. The qubits are measured in the $\sigma_x$ eigenbasis (the {$\hat{H}$} gate and the {$\sigma_z$}-basis measurement can be seen as a {$\sigma_x$}-basis measurement). The dashed line represents classical information.}
\label{fig:circ1}
\end{figure}

Consider a register of {$N$} qubits prepared in the generalized GHZ state 
\begin{equation}
\label{input}
{\ket{\Psi}_{12...N} = \frac{1}{\sqrt{2}} (\ket{0}_1 \ket{0}_2 \! \cdot \! \cdot \! \cdot \! \ket{0}_N + \ket{1}_1 \ket{1}_2 \! \cdot \! \cdot \! \cdot \! \ket{1}_N)}.
\end{equation}
(see Fig.~\ref{circ1}) The underlying assumption is that the values of the set $\{\nu_j\}$ are distributed to the {$N$} stations of a network. Each agent has knowledge of just the value that has been attributed to him. We shift the phase of the {$j$}-th qubit ({$j=1,...,N$}) by {$\frac{\nu_j}{N \theta}$} by applying the conditional operator
$\hat{R}_j (\frac{\nu_j}{N \theta})=
e^{i \frac{\nu_j}{N \theta}}\ket{0}\bra{0}+\ket{1}\bra{1}$.
After these single-qubit operations, the state of the register becomes 
\begin{equation}
{\miniket{\tilde{\Psi}}_{12...N}=\frac{1}{\sqrt{2}}(e^{i \frac{\mu}{\theta}} \ket{0}_1 \ket{0}_2 \! \cdot \! \cdot \! \cdot \! \ket{0}_N + \ket{1}_1 \ket{1}_2 \! \cdot \! \cdot \! \cdot \! \ket{1}_N)}.
\end{equation}
 By means of {$\sigma_x$}-basis measurements on each qubit but the first one, we end up with {$\miniket{\tilde{\psi}^{\pm}}_{1} = \frac{1}{\sqrt{2}}(e^{i \frac{\mu}{\theta}} \ket{0}_1 \pm \ket{1}_1)$} with the {\it plus} ({\it minus}) sign if the number of 1's measured on the other qubits is even (odd) and $\sigma_{r}$ the $r$-Pauli matrix ({$r=x,y,z$}). If the number of 1's is odd, we can obtain {$\miniket{\tilde{\psi}^{+}}_{1}$} simply by shifting the phase of the first qubit by {$\pi$}. The information about {$\mu$} is now carried by the first qubit and we can estimate it by an interference-type procedure where we apply {$\hat{H}$} to obtain {$\frac{e^{i \frac{\mu}{\theta}} + 1}{2} \ket{0}_1 + \frac{e^{i \frac{\mu}{\theta}} - 1}{2} \ket{1}_1$}. The probability of measuring {$\ket{0}$} or {$\ket{1}$} out of this state are now respectively {$\cos^2 (\frac{\mu}{2 \theta})$} and {$\sin^2 (\frac{\mu}{2 \theta})$}. Therefore, by repeating this procedure {$\alpha$} times, in virtue of the central limit theorem, we can estimate {$\frac{\modul{\mu}}{\theta}$} with a precision {${\cal O} (\frac{1}{\sqrt{\alpha}})$}. A close analogy with measurement-based computation can be seen here, as the additional phase-shift conditioned on the outcomes of the previous $N-1$ measurements represents the {\it byproduct operator} of the specific scheme at hand~\cite{cluster,iocluster}.

If we need to estimate the order of magnitude of {$\mu$}, we can start the algorithm by taking a large value of {$\theta$} (say, for example, 0.5) and evaluate the ratio {${\modul{\mu}}/{\theta}$}. If this ratio is found to be {${\cal O} (1)$}, we have the correct order of magnitude of {$\mu$}. Differently, if the ratio is much smaller than one, we divide {$\theta$} by a fixed value (say 2) and estimate again {${\modul{\mu}}/{\theta}$} until we find it to be {${\cal O} (1)$}~\cite{groveraverage}. In this way, we need {${\cal O} (\log_{2} \mu)$} applications of the quantum algorithm to solve the problem (each application of the algorithm requires {${\cal O} (\alpha)$} steps, but if we fix the precision we want in the estimate, {$\alpha$} is fixed too). On the other hand, we have seen that any classical algorithm can solve this problem in {$\Omega (\frac{1}{\mu^2})$} steps, therefore a speed-up is provided by the quantum algorithm. This protocol is particularly useful in a scenario of distributed quantum computation, where a processor is made out of a network of local nodes interconnected by classical and quantum channels~\cite{groveraverage,distributed}. Indeed, the qubits can be at remote locations and can operate independently with the partial knowledge of just the value of the $\frac{\nu_j}{N \theta}$ belonging to node $j$. The only requirement for the algorithm to work, once the initial resource is provided, is the transmission of one bit of classical information (the result of each measurement) to the first qubit location.

\subsection{Classical simulations of quantum algorithms}
\label{simulazione}
The classical wave optics analogy of quantum information processing is based on the fact that the state of a quantum system evolves according to a wave equation and satisfies the superposition principle~\cite{nielsenchuang}. Many classical simulations of quantum algorithms have been proposed in recent years. They require a number of classical resources scaling exponentially with the number of qubits being simulated~\cite{costi}. For example, a few classical optical simulations represent the Hilbert space of {$n$} qubits by considering the propagation of a classical electromagnetic wave. Splitting the cross section of this wave in {$2^n$} different spatial zones allows one to associate the amplitude of the electromagnetic wave in each zone with the amplitude of a state of the computational basis of the quantum system to be simulated~\cite{simulo,ondaclassica}. Another proposal put forward is to represent {$n$} qubits by a single photon in an interferometric setup involving {$2^n$} optical paths~\cite{cerf}. In this case the price to pay is the exponential growth of the number of optical paths and optical devices required for the implementation. Yet another way is based on the use of a single particle with {$2^n$} energy levels, where each level will embody a computational state of the register~\cite{meyer}.

All these suggestions are inherently based on the reasoning that usually a quantum algorithm consists only of transformations indiscriminately acting on all the qubits of a register (operators acting on all the states, for example {$\hat{H}^{\otimes n}$}) or specific state transformations (operators acting only on specific states, for example the phase-inversion of the searched state in the quantum search algorithm described in Sec.~\ref{grovercerca}). However, the quantum average algorithm requires operators acting on specific qubits (for example the single-qubit phase-shifts). To give a clear picture of the differences between the two kinds of transformation above and operators acting on specific qubits, we consider the phase-shift stage in the quantum average algorithm. Two possible ways to simulate  it with classical processes can be distinguished. We can consider a {\it serial} sequence of transformations, each one equivalent to a single-qubit rotation (if we simulate the Hilbert space of {$n$} qubits in the above-mentioned ways, each transformation classically corresponds to an operator acting on {$2^{n-1}$} degrees of freedom of the classical system~\cite{Kwiat}). However, we will lose the {\it parallel} computation characteristic, that is at the basis of this algorithm. The other way to classically simulate this stage is to consider a parallel application of all these transformations. But it is easy to see that, considering again the ways mentioned above to simulate the Hilbert space of the quantum system, such a parallel application reduces to a single transformation rotating one state of the classical system by an angle proportional to the average value (it also rotates the other states by different angles, but we are not interested in them). We thus need to know {\it a priori} the result of the algorithm. Therefore, the only non-trivial simulation changes radically the nature of the algorithm from {\it parallel} to {\it serial} computation. This highlights an intrinsic difference between the algorithms involving only global transformations or specific state transformations (polylocal class) and those which need transformations affecting only specific qubits (non-polylocal class).

\section{Noise effects on the representative algorithms}
\label{rumore}

Our model for noise is motivated by considerations typical of static quantum chaos, where a register is assumed to be affected by individual, time-independent imperfections on each qubit~\cite{chaos}. Here we consider the possibility of an imperfect preparation of the state of the register by allowing each qubit to be in a mixed state. Intuitively, the loss of purity of the overall state can be expected to influence the behavior of the entanglement involved, if any, in a specific quantum algorithm. We therefore consider the initial state of the {$j$}th qubit ({$j=1,..,n$}) as given by the density matrix 
\begin{equation}
\label{misto}
{\rho_j = \lambda_j \pro{0}{j}{0} + ( 1 - \lambda_j ) \pro{1}{j}{1}},
\end{equation}
where {$\lambda_j$} is the probability of finding the {$j$}th qubit in its ground state and {$\ket{0}_j$} is the ideal starting state. We name the source responsible for such an initial state as {\it static noise}. For our purposes, we do not need to identify the mechanism responsible for such imperfections to occur. This is a setup-dependent issue that will specialize our study. Nevertheless, we mention that if each {$\lambda_j$} follows the Boltzmann distribution for a two-level system, this model for mixedness represents a qubit being thermally excited. Such an assumption is not at all unrealistic: the study of quantum algorithms in the presence of non-ideal preparation can be pragmatically relevant. For instance, solid-state implementations require the cooling of a register to very low temperatures, which may be experimentally demanding and quite unnecessary if a protocol is known to perform adequately with tolerable mixed initial state. These considerations make our investigation of practical importance. We remark that this assumption of a static model for noise is only the first step toward a more complex and complete study of an important problem in quantum information science. 

\subsection{Polylocal class: Quantum search algorithm}
\label{rumoremedia}

The simulation of noise effects can be archived by choosing a set of {$\lambda_j$}'s and evaluating the probability of obtaining the searched state after each iteration of the {Grover iterate}. In order to fix the ideas and present our results in a clear way, we have considered a {\it symmetrical} case of all equal {noise parameters} $\lambda_j=\lambda,\,\forall{j} = 1,...,n$. This choice is reasonable for a spatially localized register, where all the qubits experience a noise mechanism of negligible strength fluctuations or are in touch with the same thermal environment. 

In Fig.~\ref{fig:fig3} we present the results for a system of 2, 3 and 4 qubits (panels {\bf (a)}, {\bf (b)} and {\bf (c)} respectively) where we plot the probability {$P$} of finding a searched state against the number of iterations $m$ of the {Grover iterate} and the noise parameter {$\lambda$}. 
\begin{figure}[ht]
\centerline{\bf (a)}
\psfig{figure=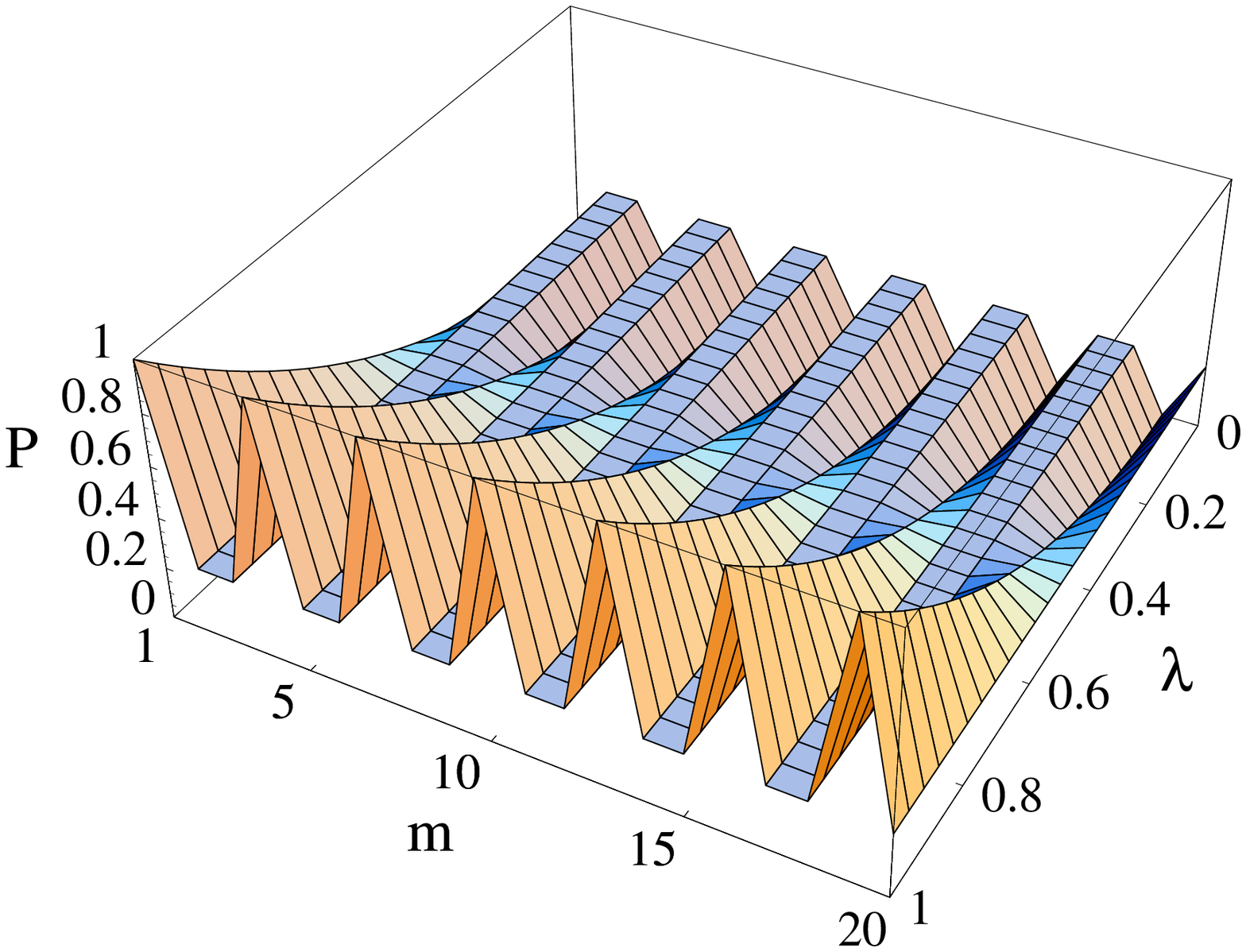,width=6cm,height=4.5cm}
\centerline{\bf (b)}
\psfig{figure=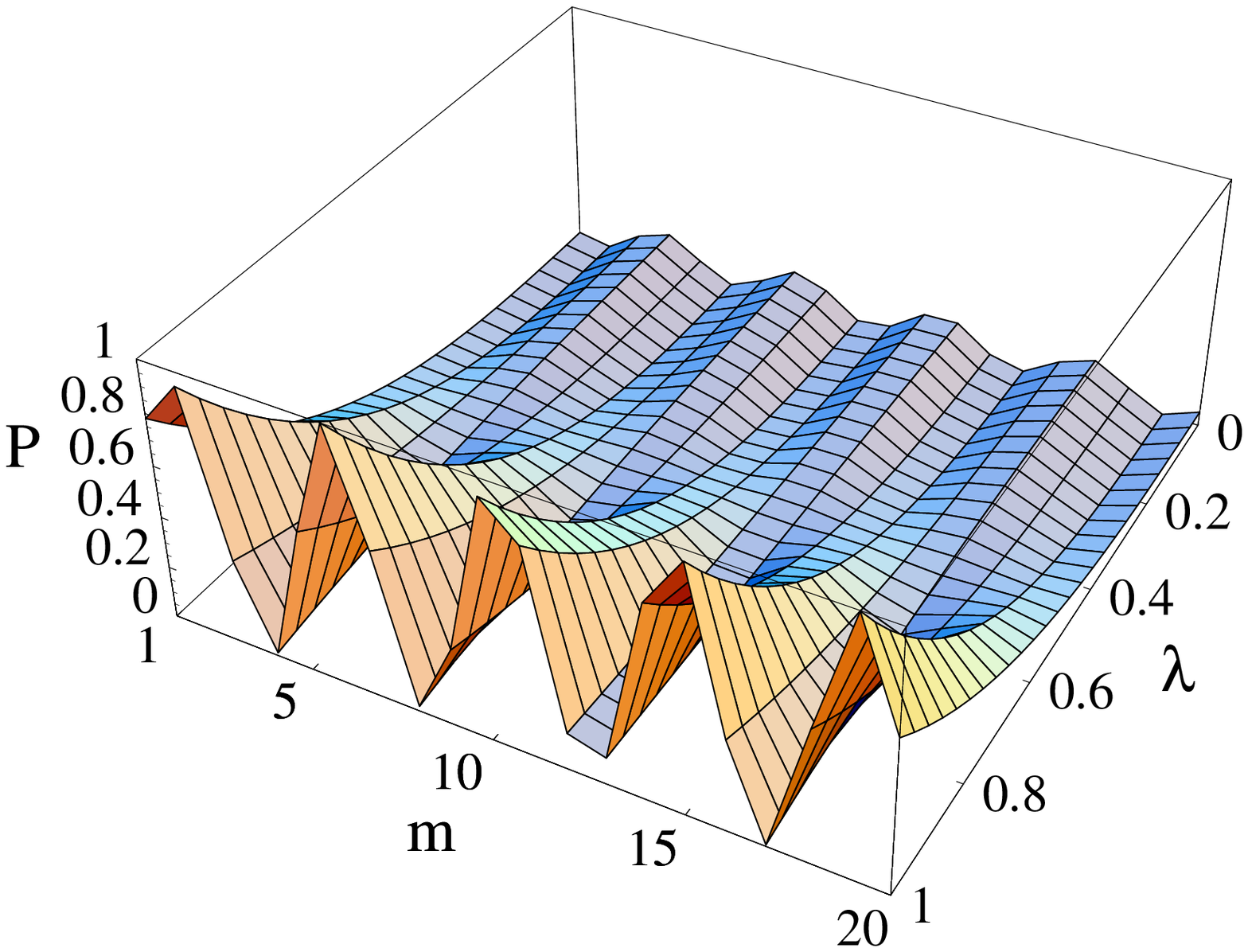,width=6cm,height=4.5cm}
\centerline{\bf (c)}
\psfig{figure=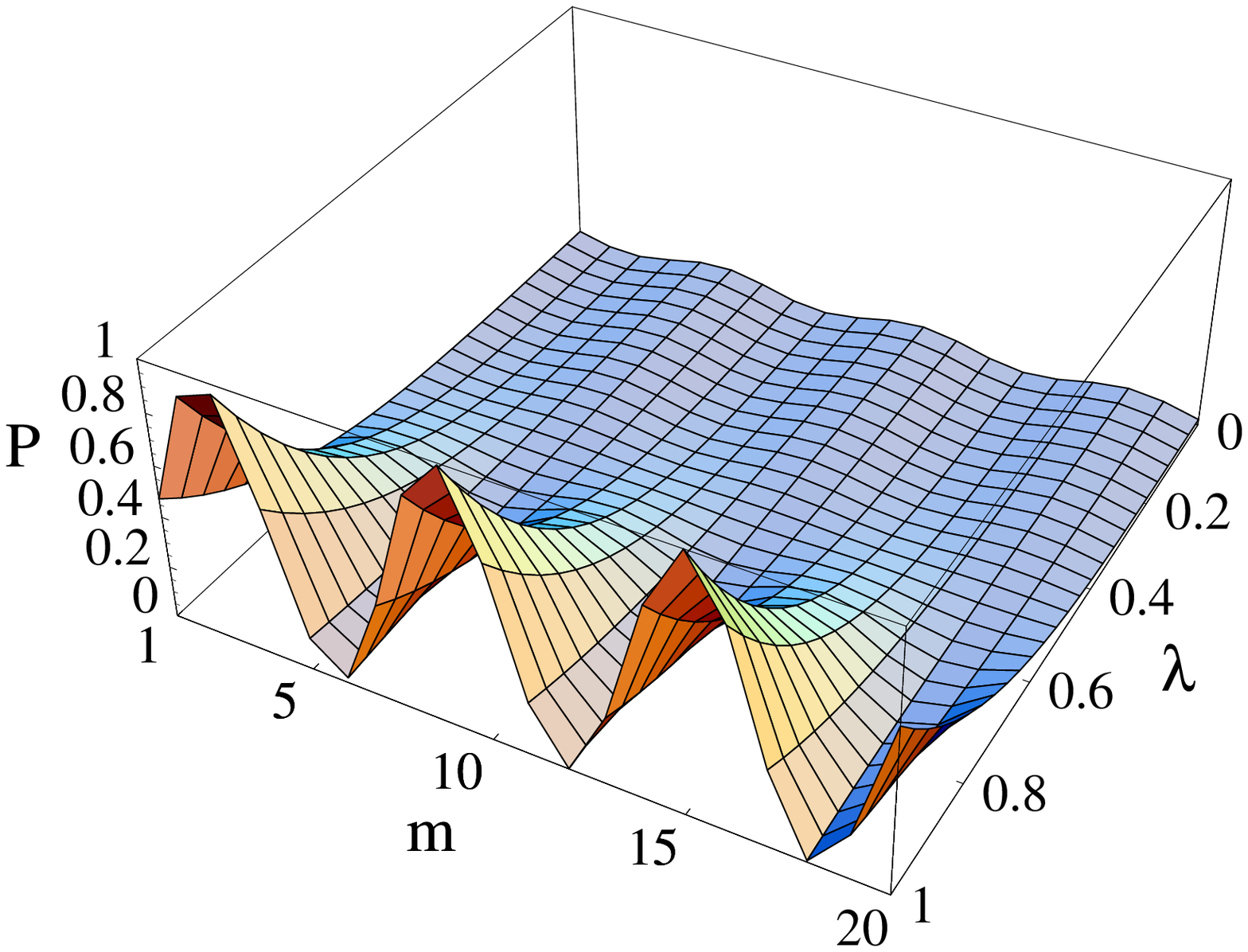,width=6cm,height=4.5cm}
\caption{(Color online) Probability {$P$} to find a searched state (regardless of its specific instance) against the number of iterations {$m$} and the noise parameter {$\lambda$} in the quantum search algorithm acting on a noise-symmetric register of 2, 3 and 4 qubits (panels {\bf (a)}, {\bf (b)} and {\bf (c)} respectively).}
\label{fig:fig3}
\end{figure}
One can clearly see that the period of the oscillations between the two extremal states involved in the algorithm (the equally weighted superposition {$\ket{\tilde{0}}$} and the searched state) is the same for any value of noise. This property makes the protocol somewhat robust against imperfections in the preparation of the register. In a situation where there is just a limited knowledge about the initial purity of the register, we do not have to make the protocol adaptive, as the period is left unchanged. This indirectly confirms the resilience of the timing of the protocol outlined in~\cite{grovernoise} for different approaches than ours. However, as the noise increases, the probability for the noisy algorithm to find the searched item is strongly affected. For any number of qubits larger than 2 and in absence of noise ({\it i.e.} for $\lambda=1$), {$P$} is not exactly 1 after the number of iterations corresponding to the first maximum (for 2 qubits it is well-known that {$P=1$} after just one iteration). We have therefore considered the {normalized probability} {$P_{norm} = {P}/{P_{ideal}}$}, where {$P_{ideal}$} is the probability to obtain the searched state when $\lambda=1$ and we have calculated {$P$} and {$P_{ideal}$} after the number of iterations corresponding to the first maximum. The results are shown in Fig.~\ref{fig:fig4} for the case of 2 ($\blacksquare$), 3 ($\blacklozenge$) and 4 qubits ($\blacktriangle$).
\begin{figure}[t]
\psfig{figure=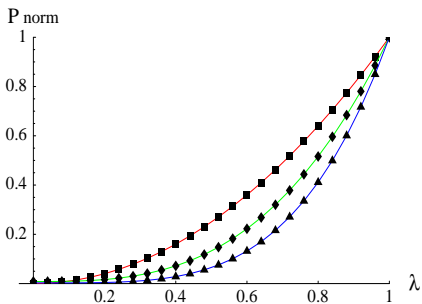,width=6cm,height=4cm}
\caption{(Color online) Normalized probability {$P_{norm}$} to find the searched state (regardless of its specific instance) against the noise parameter {$\lambda$} in the quantum search algorithm acting on a register of 2 {\bf ({$\blacksquare$})}, 3 {\bf ({$\blacklozenge$})} and 4 qubits {\bf ($\blacktriangle$)}.}
\label{fig:fig4}
\end{figure}
The plot reveals that $P_{norm}$ (which is independent of the state that has to be found) scales as {$\lambda^n$} with {$n=2,3,4$}, demonstrating a severe fragility of the scheme to static imperfections. This result can be easily generalized to an arbitrarily inhomogeneous set $\{\lambda_j\}$ and to any dimension of the register. We find that
\begin{equation}
\label{generale}
P^{\{\lambda_j\}}_{norm} \sim {\prod_j {\lambda_j}}
\end{equation}
with $P^{\{\lambda_j\}}_{norm}$ the normalized probability of obtaining the searched state in the presence of generally {\it asymmetric} noise. This result can easily be understood by closely looking at the model for imperfections. Consider for instance the two-qubit case; the initial state of the register is written as {$\rho_{12} = p_{00} \proj{00}{00} + p_{01} \proj{01}{01} + p_{10} \proj{10}{10} + p_{11} \proj{11}{11}$}.
 The probabilities {$p_{ij}$} are respectively {$p_{00} = \lambda_1 \lambda_2$}, {$p_{01} = \lambda_1 (1 - \lambda_2)$}, {$p_{10} = (1 - \lambda_1) \lambda_2$} and {$p_{11} = (1 - \lambda_1) (1 - \lambda_2)$}. Due to the linearity of quantum mechanics, we can study the evolution of each of the states involved in {$\rho_{12}$} separately. The state {$\proj{00}{00}$} is the initial state of the register in the ideal case of $\lambda_{1,2}=1$. The evolution of the system dictated by the algorithm is unitary so that the states present in the evolved state will remain orthogonal to each other. In particular, the algorithm will transform any other computational state into a state that will be orthogonal to the searched one resulting out of the evolution of {$\proj{00}{00}$}. Therefore, the probability of obtaining the searched state after the right number of iterations is precisely  {$p_{00} = \lambda_1 \lambda_2$}, as there will be no contribution from any other state. We can extend this proof to any number of qubits, therefore arriving to the result of Eq.~(\ref{generale}). Of course, for a number of qubits larger than 2, the ideal output state after the performance of the algorithm with $\lambda_j=1$ will not be precisely {$\proj{s}{s}$}. However, the contributions from the orthogonal states will be negligible. Therefore, in the case of an inhomogenous set of {$\lambda_j$}'s, results qualitatively analogous to those presented here should be expected. The fragility of the algorithm to this simple model for imperfections must be looked at in terms of the modification suffered by the entangled state ``created'' by the {Grover iterate} in the course of the protocol~\cite{braunsteinpati}.

\subsection{Non-polylocal class: Quantum average algorithm}

In order to give a full-comprehensive analysis of the effects of noise in the quantum average algorithm, we explicitly include the steps required to create the entangled resource consumed during the computation. For this purpose, we consider the control-NOT (CNOT) gate~\cite{nielsenchuang} 
\begin{equation}
\rm{CNOT}=
\begin{pmatrix} 
1&0&0&0\\ 
0&1&0&0\\
0&0&0&1\\
0&0&1&0
\end{pmatrix},
\end{equation}
written in the two-qubit basis {$\{\ket{00},\ket{01},\ket{10},\ket{11}\}$}, where the first qubit is the control and the second is the target of the gate. The GHZ-like state needed for the algorithm to work when $\lambda_j=1$ is obtained starting from the ideal initial state {$\ket{\underline{0}}$}. The application of {$\hat{H}$} to the first qubit only, followed by a set of CNOT's with the first qubit as the control and all the other qubits in the register as the targets, results in the required resource. The complete protocol is sketched in Fig.~\ref{fig:circ2}. The preparation stage of the algorithm is addressed explicitly here because the static noise affecting the register has influences on the form of the entangled state used in the computation. By incorporating these preliminary steps, we provide a more complete understanding of these influences.  
\begin{figure}[t]
\psfig{figure=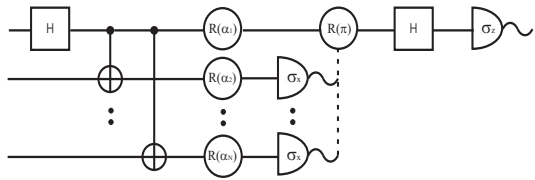,width=7cm,height=2.3cm}
\caption{Circuit for the quantum average algorithm~\cite{groveraverage} including the entangling gates for the preparation of the input state. Here, the states entering the circuit are given by the $\rho_{j}$ defined in Eq.~(\ref{misto}). CNOT gates, controlled by the first qubit in the register, are shown.}
\label{fig:circ2}
\end{figure}

The choice of a meaningful figure of merit to compare the algorithm in the absence of noise with the noisy one is not straightforward. A number of difficulties arise during a careful analysis. One might naively consider the state fidelity~\cite{nielsenchuang}
\begin{equation}
{F = {\rm Tr} (\rho_{id} \rho_{noise})}=\mbox{}_{1}\langle{\tilde{\psi}^{+}}|\hat{H}{\rho_{noise}}\hat{H}|{\tilde{\psi}^{+}}\rangle_{1}
\end{equation}
 with {$\rho_{id}=\hat{H}|{\tilde{\psi}^{+}}\rangle_{1}\langle{\tilde{\psi}^{+}}|\hat{H}$} the ideal final state of the first qubit just before its measurement and {$\rho_{noise}$} the corresponding mixed state in the presence of noise. Its expression can be obtained analytically but, even in the simple case of symmetric noise, this is too lengthy to be reported here. State fidelity of the output states is often used as a significant parameter in the evaluation of the performances of a protocol. However, in what follows, we show that considering {$F$} as a figure of merit leads to the wrong conclusions. In order to fix the idea, for the calculations we have chosen the set {$\{\nu_1,\nu_2,\nu_3\}=\{-0.775,0.25,0.675\}$} with {$\theta = 0.0625$}. This allows us to give a clear picture of our results. Obviously, other choices are equally valid. 

The application of the algorithm in the presence of noise produces the plot  shown in Fig.~\ref{fig:fig5}.
\begin{figure}[b]
\psfig{figure=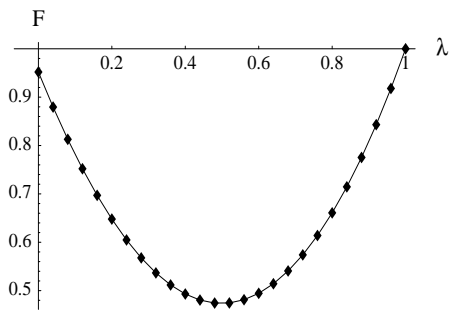,width=6cm,height=4cm}
\caption{Fidelity {$F$} of the output state in a noisy quantum average algorithm against the noise parameter {$\lambda$}, for {$\nu_1 = -0.775$}, {$\nu_2 = 0.25$}, {$\nu_3 = 0.675$} and {$\theta = 0.0625$}.}
\label{fig:fig5}
\end{figure}
For small values of {$\lambda$}, {\it i.e.} when each qubit is prepared in a state close to $\ket{1}$, $F$ is almost ideal ($F = 0.95$}). This would lead us to conclude that the protocol is effective even for a preparation {\it orthogonal} to the one designed for the algorithm to work. However, the conclusion is erroneous as revealed by immediately calculating the value {${\modul{\mu_{noise}}}/{\theta}$}, with the pedex reminding us that the algorithm has been run in the presence of noise. By assuming $\lambda = 0$ we obtain {${\modul{\mu_{noise}}}/{\theta} = 0.36$} rather than the true value {${\modul{\mu}}/{\theta} = 0.80$}, showing a considerable discrepancy. On the other hand, at {$\lambda = 0.1$} we have $F = 0.77$ (lower then the one for {$\lambda = 0$}) and an estimate {${\modul{\mu_{n}}}/{\theta} = 0.94$}, much closer to the actual value of {$\frac{\modul{\mu}}{\theta}$}. Therefore, the state fidelity would lead us to erroneously privilege the first case over the second, which actually delivers a more faithful estimate of the average. 

\begin{figure}[b]
\centerline{{\bf (a)}\hskip4cm{\bf (b)}}
\centerline{\psfig{figure=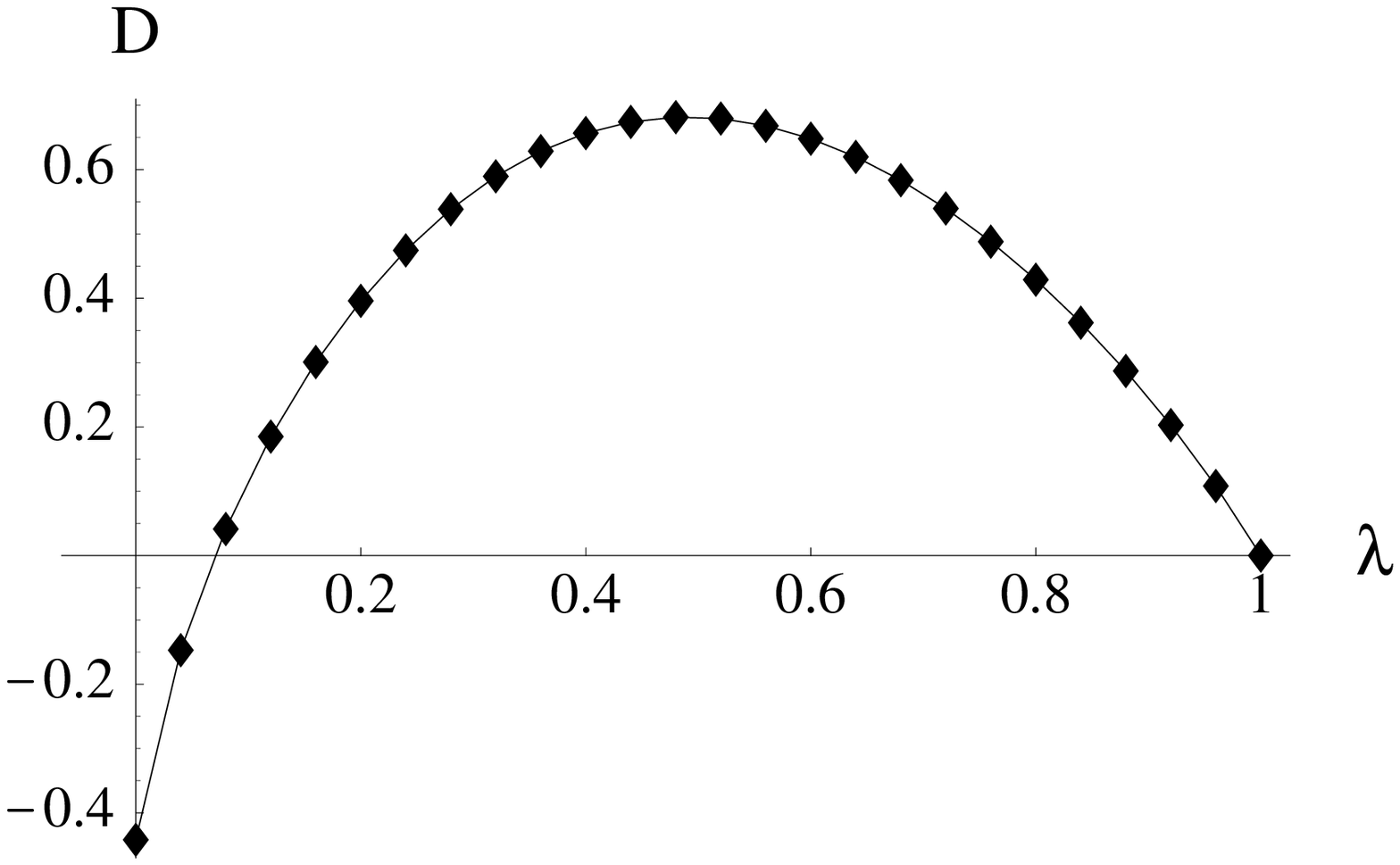,width=4.5cm,height=2.8cm}\psfig{figure=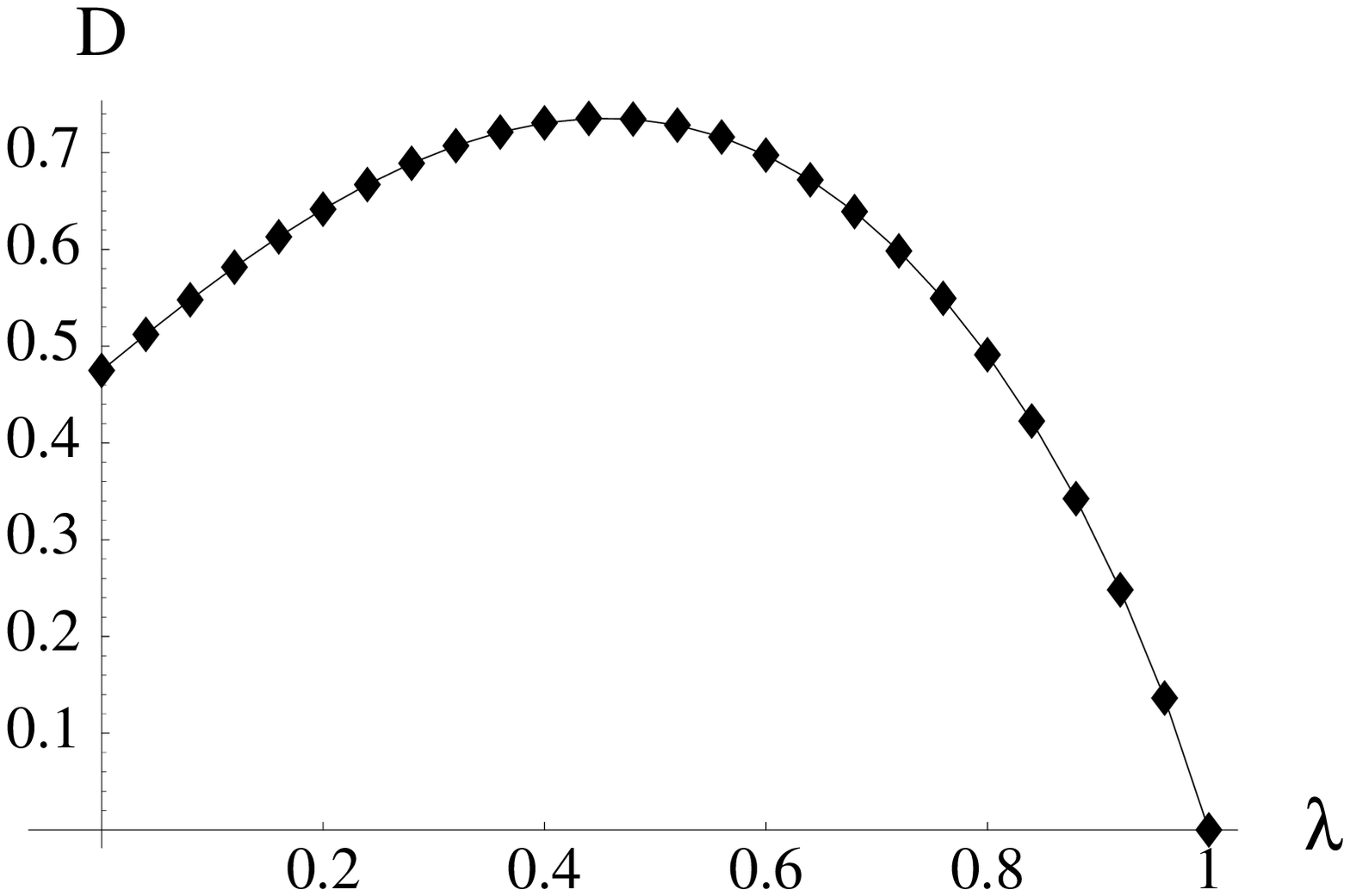,width=4.5cm,height=2.8cm}}
\caption{{\bf (a)}: Distance ratio $D$ of the output state of the quantum average algorithm against the noise parameter {$\lambda$}, for the same set of values of Fig.~\ref{fig:fig5}. {\bf (b)}: Distance ratio $D$ of the output state of the quantum average algorithm against the noise parameter {$\lambda$}, for the same set of values of Fig.~\ref{fig:fig5}, but after {$\nu_1$} and {$\nu_2$} have been swapped.}
\label{fig:fig6}
\end{figure}

A more significant performance parameter is given by the {\it distance ratio} {$D=({\modul{\mu_{noise}} - \modul{\mu}})/{\theta}$}, which measures the distance between the true and estimated average in units of $\theta$.  Obviously, $D\simeq{0}$ implies that the noise does not spoil the accuracy of the computation. By using $D$ in the same situation considered above, we obtain much more faithful information, as shown in Fig.~\ref{fig:fig6} {\bf (a)}, 
where we can notice that, for {$\lambda\simeq{0.1}$}, the algorithm gives a reliable evaluation of the ratio {${\modul{\mu}}/{\theta}$}. This might seem surprising at first sight. However, it is easy to recognize that this is simply a fortuitous case due to the dependence of the algorithm on the actual set of $\nu_j$'s. Indeed, suppose we swap the value of {$\nu_1$} and {$\nu_2$}: obviously the average value remains unchanged. However, we obtain a different plot for $D$ (see Fig.~\ref{fig:fig6} {\bf (b)}). This is not a feature of the chosen figure of merit but an intrinsic characteristic of the quantum average algorithm. In a different way to the search protocol, here the qubits in the register play unequal roles. Indeed, in addition to carrying information about the first element of the set $\{\nu_j\}$ after the phase-shift stage, the first qubit is also responsible for the information about the average value after the measurements on all the other qubits~\cite{commentofedelta}. In practice, this may represent a problem: in a noise-asymmetrical setting, the noise affecting the last qubit to be measured is critical in determining the ``quality'' of the evaluated average. Ideally one would like to screen it from noise in order to have a more faithful estimate.

A possible way to circumvent this problem is to consider a variation of the algorithm in which an enlarged register of $N+1$ elements is used. The first qubit is measured at the end of the algorithm while the rotations are performed on all the other {$N$} qubits. We assume that the first qubit is protected and in a pure state, while all the others are prepared in {$\rho_j$}'s. This situation is reminiscent of analogous investigations performed with respect to a different quantum algorithm~\cite{plenioshor}. Moreover, this scheme resembles the paradigm used in the model for {\it deterministic quantum computation with one quantum bit}~\cite{dqc1}. Both the above cases showed that one pure state qubit singled out from a register prepared in a statistical mixture is sufficient to carry out several computational protocols. Our study reinforces such ideas and at the same time suggests an operative way to limit the effects of static imperfections. It is important to stress that by shielding the ruler qubit in such modified protocol, we want to effectively avoid the accumulation of noise effects rather than fix the mistakes occurred during the computation. This is different, in both motivations and strategy, from quantum error correction techniques~\cite{nielsenchuang,preskill}. No redundant encodings is introduced, in our scheme, which are instead typical of error correction protocols. Each qubit in our modified scheme is a physical information carrier rather than a logical one encoded into the Hilbert space of a block of qubits.  

\begin{figure}[t]
\psfig{figure=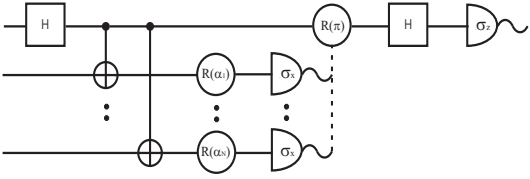,width=7cm,height=2.3cm}
\caption{Circuit for the modified quantum average algorithm with the introduction of the {\it ruler} qubit, which is prepared in a pure state that is not phase-shifted in the course of the protocol. All the remaining qubits enter the algorithm in $\rho_j\,(j=2,...,N+1)$.}
\label{fig:circ3}
\end{figure}

The logical circuit of the modified algorithm is shown in Fig.~\ref{fig:circ3}, where one can see that the required phase-shifts are now performed on all the qubits except the first in a register of $N+1$ elements. The first qubit, also known as {\it the ruler}, has to be physically distinct with respect to all the others. This is in line with a scenario of distributed computation, where the quantum average algorithm was conceived: The register configuration can be that of a star graph with the ruler at the center and all the remaining qubits occupying the outer vertices. Each qubit is connected to the ruler by classical and quantum channels, needed to exchange the information acquired after the measurements and construct the entangled resource. 

\begin{figure}[t]
\psfig{figure=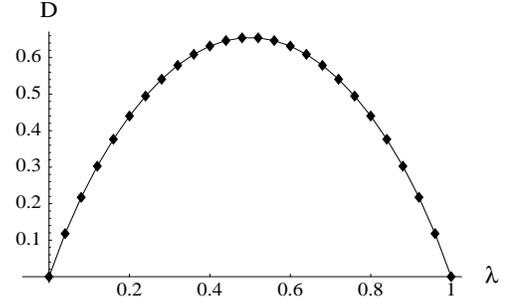,width=7cm,height=4cm}
\caption{Distance ratio $D$ of the modified quantum average algorithm with a ruler qubit against {$\lambda$} and for the same set of values of Fig.~\ref{fig:fig5}.}
\label{fig:fig8}
\end{figure}

Using the same set {$\{\nu_j\}$} as before, we obtain the behavior of $D$ shown in Fig.~\ref{fig:fig8}.
For {$\lambda = 0$} we have {$\frac{\modul{\mu_{noise}}}{\theta}=\frac{\modul{\mu}}{\theta}$} so that for this value of {$\lambda$} as well the algorithm works perfectly. Indeed, right before the application of the Hadamard and CNOT gates, the register is in the pure state {$\ket{0}_1 \ket {1}_2 \cdot \cdot \cdot \ket{1}_{N+1}$}, which then takes the form 
\begin{equation}
\ket{\Psi'}_{1..N+1}={\frac{1}{\sqrt{2}} (\ket{0}_1 \ket{1}_2 \cdot \cdot\ket{1}_{N+1} + \ket{1}_1 \ket{0}_2 \cdot \cdot \ket{0}_{N+1})}.
\end{equation}
This is still a GHZ-like state and thus of the correct entanglement structure (formally $\ket{\Psi'}_{1...N+1}=\sigma_{x1}\ket{\Psi}_{1..N+1}$). By performing the algorithm with $\ket{\Psi'}$ we obtain the same probability {$\cos^2 (\frac{\mu}{2 \theta})$} ({$\sin^2 (\frac{\mu_{}}{2 \theta})$}) that the ruler qubit is in {$\ket{0}_1$} ({$\ket{1}_1$}). 
The dependence on the order of the values has disappeared, so that the use of the distance ratio in a modified protocol with a ruler qubit now allows us to perform a faithful assessment of a noisy algorithm. 

However, a further problem to address is the dependence of $D$ on the average value. Here, we study the maximum amount of noise that the algorithm can tolerate without affecting the ratio {${\modul{\mu}}/{\theta}$}, independently of the set {$\nu_j$} (and therefore of the average value) and {$\theta$}. 
As our task is to check that {${\modul{\mu}}/{\theta}$} is {${\cal O} ( 1 )$}, a precision of {$0.5$} will be considered as acceptable. As the behavior of $D$ will now be symmetrical with respect to {$\lambda = \frac{1}{2}$} (corresponding to $\rho_j$'s being completely mixed), we decide to use the {\it purity parameter} {$\tau = \modul{2 \lambda - 1}$} to quantify the strength of the imperfections. Obviously, for {$\tau = 0$} the noise will be maximum (all the input qubits will be completely mixed) while for {$\tau = 1$} the input qubits will be in the pure state {$\ket{0}_j$} or {$\ket{1}_j$} (in which case the ratio {${\modul{\mu}}/{\theta}$} will be correctly evaluated, as we have shown before). The probability of obtaining {$\ket{0}_1$} in the presence of noise is evaluated to be {$P_N (\tau) = \arccos [ \sum_{i = 0}^{\rm{Int}( \frac{N}{2} )} (-1)^i \tau^{2 i} A_{( N - 2 i )}^{( 2 i )} ]$} with
\begin{equation}
\rm{Int}\left(\frac{N}{2}\right)=\left\{
\begin{aligned} 
&\frac{N}{2}\;\;\;\;\;\;\;\;\;\;\;\;\;\;\;\rm{for}\;\rm{even}\;N\\  
&\frac{N-1}{2}\;\;\;\;\;\;\;\;\;\rm{for}\;\rm{odd}\;N
\end{aligned}
\right.
\end{equation}
and 
\begin{equation}
A_{(m)}^{(l)}\!=\!\sum_{perm}{\cal P}\left[\sin\left(\frac{\nu_{j_1}}{\theta}\right) \! \cdot \! \cdot \! \cdot \sin \left(\frac{\nu_{j_l}}{\theta}\right) \cos\left(\frac{\nu_{k_1}}{\theta}\right) \! \cdot \! \cdot \! \cdot \! \left(\frac{\nu_{k_m}}{\theta}\right)\right].
\end{equation}
Here, {${\cal P}$} is the permutation operator for the indices {$j$} and {$k$}, which are in number of {$l$} and {$m$} respectively. For example
\begin{equation}
\begin{aligned}
A_{( 1 )}^{( 2 )}&=\sin \left(\frac{\nu_1}{\theta}\right)\sin \left(\frac{\nu_2}{\theta}\right) \cos \left(\frac{\nu_3}{\theta}\right)+\\
&+\sin \left(\frac{\nu_1}{\theta}\right) \cos \left(\frac{\nu_2}{\theta}\right) \sin \left(\frac{\nu_3}{\theta}\right)+\\
&+\cos \left(\frac{\nu_1}{\theta}\right) \sin \left(\frac{\nu_2}{\theta}\right) \sin \left(\frac{\nu_3}{\theta}\right).
\end{aligned}
\end{equation}
It is straightforward to notice that {$P_N(1)=\cos^2 (\frac{\mu}{2 \theta})$}.

Rather than find the values of {${\nu_j}/{\theta}$} that maximizes {$\modul{P_N (\tau) - P_N (1)}=\modul{{\modul{\mu_{noise}} - \modul{\mu_{id}}}}/{\theta}$} for a set value of {$N$}, we can maximize the absolute value of the difference of the arguments in the inverse cosine as this is a monotonic function in the interesting range of values. We find that {$| \sum_{i = 0}^{\rm{Int} ( \frac{N}{2} )} (-1)^i \tau^{2 i} A_{( N - 2 i )}^{( 2 i )} - \sum_{i = 0}^{\rm{Int} ( \frac{N}{2} )} (-1)^i A_{( N - 2 i )}^{( 2 i )} |$} is maximum for {${\nu_j}/{\theta}$}'s all equal to a {$\tilde{\nu}_{max}^{(N)}$} depending only on the number of qubits but otherwise independent of {$\tau$}. We have numerically calculated the value of {$\tilde{\nu}_{max}^{(N)}$} for {$N=3,...,8$ and report in Fig.~\ref{fig:fig9} the corresponding {$\modul{P_N (\tau) - P_N (1)}$}.
 
\begin{figure}[t]
\psfig{figure=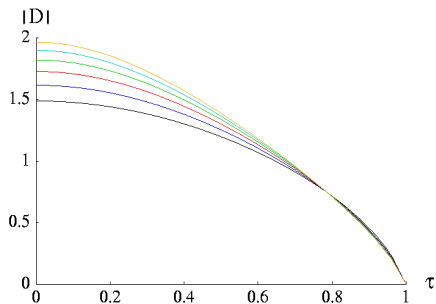,width=6cm,height=4cm}
\caption{(Color online) Maximum value of $|D|$ against the purity parameter {$\tau$} for a register of 3 to 8 elements. In this plot, the number of qubits diminishes by one (starting from $N=8$) in going from the top curve to the bottom one.}
\label{fig:fig9}
\end{figure}
The noise that the algorithm can tolerate while still giving a faithful estimate of {${\modul{\mu}}/{\theta}$} with a precision of {$0.5$}, regardless of the values of {${\nu_j}/{\theta}$}, is {$\tau\simeq{0}.90 $} and slightly depends on the number of qubits. The reason for this noticeable resilience to static imperfections is explained by analyzing the global density matrix of the register during the performance of the algorithm. We give an account of this robustness in the next section, which sheds light on the behavior of entanglement in the protocol itself.

\section{Analysis of entanglement in the quantum average algorithm}
\label{entanglement}

The study of the entanglement behavior in the presence of noise can help to understand the reasons for the resilience we have highlighted above. The fragility of the GHZ-like resource consumed in the algorithm may lead us to think that even simple static imperfections in the register would have dramatical effects on the quality of the computation.

In analogy with what was done in the previous section, we consider the modified version of the quantum average algorithm with a ruler qubit. As the conditional phase-shifts used are local unitary operations, the degree of entanglement in the system is not modified by their application. Therefore, we can study the entanglement in the register before their action (of course the measurements performed at the end of the protocol will consume the entanglement). It is important to note that the entanglement dynamics are in contrast to those responsible for the speed-up in polylocal algorithms. In those cases, entanglement has to be created during the protocol and spread all over the register in order for the algorithms to outperform their classical analogs~\cite{jozsa,vidal1, braunsteinpati}. For the quantum average algorithm, a set of measurements progressively breaks preconstituted quantum correlations.

The focus of our interest is bipartite entanglement in the system. For this analysis we consider the {$( N + 1 )$}-qubit system as split into two subgroups. We reveal entanglement between bipartitions using the Peres-Horodecki negativity of partial transposition criterion~\cite{pereshorodecki}. Even though this test is not necessary and sufficient for revealing quantum correlations in the case of a general multipartite register, it is a useful tool for the present analysis. We first examine a three-qubit system and then generalize the results to any number of qubits. 

When dealing with three qubits, one can either consider the bipartite entanglement in the reduced state obtained by tracing out the degrees of freedom of one of the qubits (we call it the {\it traced case}) or look for the correlations between one qubit and the remaining two, considering every possible permutation of qubit labels (we refer to this as the {\it non-traced case}). In the traced case, we have considered both the trace with respect to the ruler qubit (therefore studying the presence of bipartite entanglement between two mixed qubits) and the trace with respect to one of the mixed qubits of the register (thus evaluating the quantum correlations of the remaining register qubit with the ruler). The result is that in both cases no bipartite entanglement is present. By studying the non-traced case, regardless of the configuration of the bipartitions, we have obtained that {$\forall \tau \ne 0$} bipartite entanglement is present. These results are a reminder of the properties of GHZ states. General considerations for any number of qubits can be obtained by analyzing the static noise we have chosen. With this model, the density matrix of the whole system, before the phase-shift stage, can be seen as an ensemble of density matrices of GHZ-like form. For instance, the global density matrix for the three-qubit system in the presence of noise is 
\begin{equation}
\begin{aligned}
\rho& = \lambda_1 \lambda_2 \miniproj{GHZ_{00}^{(2)}}{GHZ_{00}^{(2)}}\\
& + \lambda_1 ( 1 - \lambda_2 ) \miniproj{GHZ_{01}^{(2)}}{GHZ_{01}^{(2)}}\\
& + ( 1 - \lambda_1 ) \lambda_2 \miniproj{GHZ_{10}^{(2)}}{GHZ_{10}^{(2)}}\\
& + ( 1 - \lambda_1 ) ( 1 - \lambda_2 ) \miniproj{GHZ_{11}^{(2)}}{GHZ_{11}^{(2)}}
\end{aligned}
\end{equation}
where we have introduced the set of generalized GHZ-like states
\begin{equation}
\begin{aligned}
&|{GHZ_{00}^{(2)}}\rangle = \frac{1}{\sqrt{2}} ( \ket{000} + \ket{111} ),\\
&|{GHZ_{01}^{(2)}}\rangle = \frac{1}{\sqrt{2}} ( \ket{001} + \ket{110} ),\\
&|{GHZ_{10}^{(2)}}\rangle = \frac{1}{\sqrt{2}} ( \ket{010} + \ket{101} ),\\
&|{GHZ_{11}^{(2)}}\rangle = \frac{1}{\sqrt{2}} ( \ket{011} + \ket{100} ).
\end{aligned}
\end{equation}
In general, for an {$N$}-qubit system, the global density matrix will be the sum of projectors {$\miniproj{GHZ_{ \{ a_i \} }^{(N)}}{GHZ_{ \{ a_i \} }^{(N)}}$} with 
\begin{equation}
{\miniket{GHZ_{ \{ a_i \} }^{(N)}} = \frac{1}{\sqrt{2}} ( \ket{ 0 \; a_1 \! \cdot \! \cdot \! \cdot a_N  } + \ket{ 1 ( 1 - a_1 ) \! \cdot \! \cdot \! \cdot \! (1 - a_N ) } )}
\end{equation}
 and {$ \{ a_i\}$} the ordered sequence of digits of a binary number between {$0$} and {$2^N - 1$}. The coefficent of each {$\miniket{GHZ_{ \{ a_i \} }^{(N)}}$} is {$C_{ \{ a_i \} }^{(N)} = \prod_i \lambda_i^{a_i} ( 1 - \lambda_i ) ^{ 1 - a_i } $}. When dealing with an initial state {$\miniket{GHZ_{ \{ a_i \} }^{(N)}}$}, the protocol gives an estimate of the absolute value of a modified average {$\tilde{\mu} = \frac{1}{N} \sum_{j = 1}^N (-1)^{a_j} \nu_j$}. In presence of a not-too-severe noise, the dominating coefficents {$C_{ \{ a_i \} }^{(N)}$} are those with fewer {$1$}'s in the set {$ \{ a_i \} $}. The corresponding estimate of the average will be close to the actual one. On the other hand, the errors that result from the states with a large estimate discrepancy (those with a number of {$1$}'s in the set {$ \{ a_i \} $} close to {${N}/{2}$}) will be damped by the corresponding small coefficents {$C_{ \{ a_i \} }^{(N)}$}. This explains the reason why the algorithm is robust. This same reasoning can be applied to explain the entanglement behavior of the register during the performance of the algorithm. Indeed, the state of the register can be seen as an ensemble of GHZ-like states that will maintain the corresponding characteristics even for large values of $\lambda$.

We have also analyzed the algorithm under the effects of a different model of static noise. Assuming that the initial state of the register is
\begin{equation}
 \rho = \tilde{\tau} \; \miniproj{GHZ_{ 0 0 \cdot \cdot \cdot 0 }^{(N)}}{GHZ_{ 0 0 \cdot \cdot \cdot 0 }^{(N)}} + \frac{ 1 - \tilde{\tau}}{N} \; \mathbb{I}
\end{equation}
({\it i.e.} we are now considering white noise) the algorithm becomes fragile also for $\tilde\tau\simeq{1}$ as the state of the register can no longer be seen as an ensemble of GHZ-like states. This is further proof that the GHZ-like nature of entanglement is a pre-requisite for the efficiency of this specific algorithm.

\section{Remarks}
\label{conclusioni}
We have studied a representative of the non-polylocal class of algorithms (in contrast with the polylocal class considered by Meyer \cite{meyer}). We have found that a model of static noise preserving the GHZ-like nature of the entanglement in the register has no dramatic effects on the algorithm. On the other hand, if this specific entanglement structure is lost, the efficiency of the algorithm in term of accuracy is compromised. These results imply that the GHZ-like nature of the entanglement has a fundamental role in this specific protocol. The dependence on other classes of entanglement in other non-polylocal algorithms deserves further investigation and will be the subject of future study. 

\acknowledgments 

We thank Professor G. M. Palma and M. S. Tame for discussions and encouragement. We acknowledge financial support from the UK EPSRC. MP is supported by The Leverhulme Trust (ECF/40157).


\begin{thebibliography}{99}
\bibitem{commentonotazione} In this paper we use the following asymptotic formalism, which is typical in computer science: We say that a function $f(n)$ is {${\cal O} ( g (n) )$} [{$\Omega ( g (n) )$}] if there are two positive constants {$c$} and {$n_0$} such that, for any {$n \ge n_0$} we have {$0 \le f (n) \le c \, g (n)$} [{$0 \le g (n) \le c \, f (n)$}]. The function {$f (n)$} is said to be {$\Theta ( g (n) )$} {\it iff} $f(n)$ is ${\cal O}(g(n))$ and ${\Omega}(g(n))$ at the same time.
\bibitem{commentofedelta} The distinctive nature of the first qubit in the quantum average algorithm is also exemplified by the explicit calculation of the state fidelity $F$ discussed in Sec.~\ref{rumoremedia}. Indeed, it turns out that, in the noise symmetric setting, $F$ does not depend on $\nu_1$ for any value of $\lambda$.
\end{thebibliography}
\end{document}